\documentclass[prl,nofootinbib,twocolumn,superscriptaddress,preprintnumbers,balancelastpage,longbibliography]{revtex4-1}

\usepackage{hyperref}
\usepackage{graphicx}
\usepackage{color}
\usepackage{hepunits}
\usepackage{amssymb}
\usepackage{cleveref}
\usepackage{cancel}
\usepackage{mathrsfs}
\usepackage[normalem]{ulem}
\usepackage[english]{babel}

\usepackage[T1]{fontenc}    
\usepackage{calligra}
\DeclareMathAlphabet{\mathcalligra}{T1}{calligra}{m}{n}

\newcommand{\be}{\begin{equation}}
\newcommand{\ee}{\end{equation}}

\newcommand{\GeV}{{\, {\rm GeV}}}

\newcommand{\mpl}{M_{\rm pl}}

\makeatletter
\AtBeginDocument{%
  \def\@hangfrom@section#1#2#3{\@hangfrom{#1#2}#3}%
  \def\@hangfroms@section#1#2{#1#2}%

  \def\section{%
    \@startsection{section}{1}{\z@}%
    {0.8cm plus 0.3cm minus 0.2cm}%
    {0.5cm plus 0.1cm minus 0.1cm}%
    {\normalfont\normalsize\bfseries\centering}%
  }%

}
\makeatother

\begin{document}
\title{Axiverse Strings Resolved}

\author{Rudin Petrossian-Byrne}
\affiliation{SISSA, Via Bonomea 265, Trieste 34136, Italy}
\affiliation{INFN, Sezione di Trieste, Via Valerio 2, I-34127 Trieste, Italy}

\author{Giovanni Villadoro}
\affiliation{Abdus Salam International Centre for Theoretical Physics, Strada Costiera 11, 34151, Trieste, Italy}
\affiliation{INFN, Sezione di Trieste, Via Valerio 2, I-34127 Trieste, Italy}

 


\begin{abstract}


Axiverse cosmic strings are resolved by non-perturbative string states with large tension. We show how, in some regions of moduli space, they dissolve into low-tension configurations that are fully captured by field-theoretic solitonic solutions, such as ’t Hooft–Polyakov-like strings, allowing a post-inflationary cosmology within the original axiverse paradigm.



\end{abstract}

\maketitle

\section{Introduction}
\label{sec:Intro}

It is well known that axions can descend from higher-dimensional Abelian gauge fields. An attractive feature of such axions is that their masses can easily be exponentially small, a property often referred to as ``high quality.'' This occurs as long as the scale $\Lambda$ associated with possible shift-symmetry-breaking effects induced by charged states lies above the compactification scale $1/L$.
While the mechanisms are fully graspable in field theory with extra dimensions (see e.g.~\cite{Choi:2003wr}), the underlying motivation is best supported by the so-called string theory `axiverse' \cite{Arvanitaki:2009fg}, wherein numerous such axions may be expected \cite{Svrcek:2006yi,Acharya:2010zx,Cicoli:2012sz}.

In recent years, some activity has gone into understanding cosmic string solutions corresponding to these axions \cite{March-Russell:2021zfq,Benabou:2023npn,Reece:2024wrn}. From a phenomenological perspective, an important question is whether axiverse axions can realise the so-called ``post-inflationary'' cosmological scenario, characterized by the efficient production, evolution, and eventual decay of a cosmic string network, leading to a rich and predictive set of signals (see e.g. \cite{Agrawal:2019lkr,Gorghetto:2020qws,Gorghetto:2021fsn,Gorghetto:2022ikz,Saikawa:2024bta,Kim:2024wku,Benabou:2024msj,Gorghetto:2024vnp,Gorghetto:2025uls} for recent developments).
This compatibility was recently successfully addressed in the open string sector of the axiverse \cite{Petrossian-Byrne:2025mto,Loladze:2025uvf}.
Axion strings from the original, closed string sector of the axiverse are instead typically identified with non-perturbative high-tension stringy objects, 
such as NS/D-brane configurations, whose efficient production in a reliably calculable cosmic history poses a challenge.
What is essentially missing is a purely field-theoretic resolution, and this work aims to provide one.

We start by emphasizing that axion string solutions are necessarily magnetic sources for
the extra-dimensional gauge field from which they descend.
It is well known that magnetic charges can be resolved by the restoration of non-Abelian gauge symmetry.
We then argue in detail how a 't Hooft-Polyakov magnetic solution in higher dimensions can project down to an axion string in 4D. We describe the necessary ingredients for this construction to implement the QCD axion, more general axion-photon couplings, and a calculable post-inflationary cosmological scenario.  
Finally, we show how, in string theory, such solutions describe axiverse cosmic strings in some regions of moduli space
where the original NS/D-brane configurations `dissolve' into low-tension field-theoretic solitons.

\section{Axion strings are magnetic charges}

We will focus first on the simplest case of a $U(1)$ gauge field $A_M$ in 5D, where $M=\mu,5$. 
When compactified on a circle $S^1$ of length $L$, this descends to a $U(1)$ gauge theory in 4D, plus an axion $a(x^\mu)$. 
This axion is best identified with the gauge-invariant Wilson loop  $e^{ia/f} \equiv  e^{ i g_5 \oint_{S^1} dx^5 A_5}$, with $g_5$ the 5D gauge coupling normalised so that the smallest charge is unity. From this, its fundamental periodic domain is easily read off. The decay constant $f = 1/(g_4L)$ is controlled by the KK scale $ L^{-1}$ and the $4D$ gauge coupling $g_4 = g_5/\sqrt{L}$. 

\begin{figure}[t]
        \centering
         \includegraphics[width=0.3\textwidth]{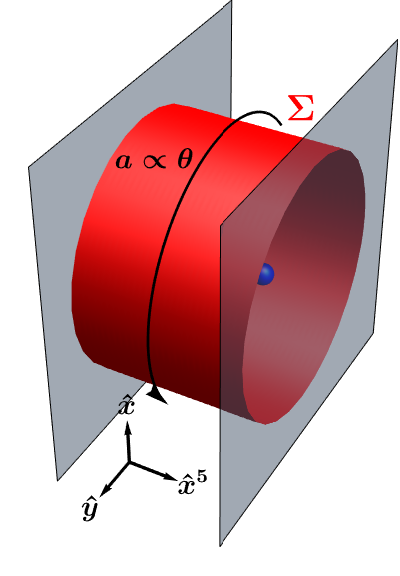}
         %
        \caption{Non-trivial winding of the axion field in the $x-y$ plane is equivalent, when resolving the compact extra dimension, to net magnetic flux through the Gaussian surface $\Sigma$ in red, revealing the presence of magnetic charge (blue dot).
        }
        %
        \label{fig:Gaussian}
\end{figure}

An axion string is  a field configuration characterized by the axion winding its fundamental domain as one traces a loop in space.
Concretely, we can take the winding in the $x-y$ plane with $\theta$ the angle therein with respect to $\hat x$. We imagine the string extended along the $z$ axis, which, as a direction of symmetry, will be a spectator and ignored for most purposes. We may thus restrict attention to $r^i=(x,y,x^5)$ space, where $x^5\in [-{L/ 2},{L/ 2}]$ 
with end points identified. 
Starting from the definition of the axion winding charge $n$ in 4D, and using the definition of the axion in terms of $A_5$, it is straightforward to show that
\begin{align}
\label{eq:Winding=MagneticCharge}
        2 \pi n &= \oint d\theta \,  {\partial_\theta a\over f}  = g_5\int_\Sigma d\theta dx^5 \, F_{\theta 5}
       \ ,
\end{align}
where the surface $\Sigma$ can be visualized in \cref{fig:Gaussian}. 
Cutting out the middle man, \cref{eq:Winding=MagneticCharge} is a statement of Gauss's law for magnetism, relating the magnetic flux through $\Sigma$ to magnetic charge enclosed, which is an integer multiple of $2\pi/g_5$, consistent with Dirac quantization.
Thus, for an axion descending from an extra-dimensional gauge field, \textit{axion string solutions are necessarily magnetic sources of the underlying gauge theory}.


Magnetic charges in 5D are intrinsically string-like objects, as minimally charged under the 2-form dual to the gauge potential $A_M$. 
In $\mathbb{R}^3\times S^1$, these can be arranged in two topologically different ways. 
The magnetic string can be either wrapped around the compact $x^5$ dimension to form a closed loop, or it can be aligned along a non-compact direction (e.g. the $z$-axis). 
The first configuration will project down to a magnetic monopole in 4D, while the second instead to an axion string. 

In the Dirac limit of point-like sources,
the appropriate gauge configuration can be constructed explicitly by method of images. 
A Dirac monopole in $\mathbb{R}^2\times S^1$ is equivalent to an infinite chain of monopoles in $\mathbb{R}^3$ with period $L$. Thus, by linearity, a valid choice centered at $r^i=0$ is
\begin{align}
\label{eq:DiracAxionString}
    \vec{A}^D = \sum_{k=-\infty}^{\infty} {g_m \over 4 \pi r_k(r_k+x)}\left[-(x^5-k L)\,\hat{y}  +y\, \hat{x}^5 \right]
\end{align}
where $r_k\equiv \sqrt{x^2+y^2+(x^5-kL)^2}$, and $g_m = n 2\pi  /g_5$ is the magnetic charge. The singular Dirac string (a sheet when restoring the $z-$axis) lies along the negative $x$ axis. This gauge artifact disappears when considering gauge-invariant quantities, as is the axion itself: 
\begin{align}
\label{eq:aOverfDirac}
    {a \over f} &= g\int_0^L dx^5 A^D_5 
   = \int_{-\infty}^\infty  dx^5 \, {gg_m y\over 4 \pi r_0(r_0-x)} = n\theta \ ,
\end{align}
which simply restates the one-to-one relation between axion winding and magnetic charge. Eq.~(\ref{eq:aOverfDirac}) is of course singular at $x=y=0$, where the Dirac charge sits. Fully resolving the core of the axion string is the same exercise of resolving magnetic singularities. Our next aim is to obtain this within the 5D EFT.

\section{Field-theoretic resolution}
\label{sec:resolution}

It was shown long ago by 't Hooft \cite{tHooft:1974kcl} and Polyakov \cite{Polyakov:1974ek} that non-singular magnetic monopoles exist in field theory  as topologically protected solitons when the $U(1)$ derives from a spontaneously broken non-Abelian gauge theory. An exact, analytic solution exists in the BPS limit \cite{Prasad:1975kr}. A magnetic string in 5D can be resolved by a trivial extension. One simply takes the 4D solution for the gauge fields $A^a_\mu$ and symmetry-breaking scalars, and extends them symmetrically into the extra dimension, while taking the extra components of $A_M^a$ to be zero.

Compactifying one dimension 
will not affect the existence of magnetic-charge solutions to the classical field equations, though it will deform the configuration. As before, the desired solution on $\mathbb{R}^2 \times S^1$ is equivalent to an infinite chain of 't Hooft-Polyakov monopoles with period $L$, but the non-linearity of the equations of motion now precludes a treatment by images. The existence of `periodic monopole' solutions has been discussed in the BPS case, first in ref.~\cite{Cherkis:2000cj}, with some further studies \cite{Ward:2005nn,Dunne:2005pr,Maldonado:2012nw}. For our purposes, the phenomenologically relevant limit (leading to a high-quality axion in 4D)  will correspond to the resolution occurring at distances much smaller than the size of the extra dimension $\ell_{\rm core} \ll L$, so the global topology of space will only affect the far-field behavior and not the core. At distances larger than $\ell_{\rm core}$ one can match, with high precision, onto the Dirac case above.

The minimal construction involves an $SU(2)$ broken to the axion-containing $U(1)_{\rm ax}$ by the expectation value of a `Higgs' $\Phi_a$ in the adjoint representation. We take $\langle \Phi_a\Phi_a \rangle =v^3$ from a potential $V(\Phi)=\lambda \left(\Phi_a\Phi_a - v^3\right)^2$. While the vacuum can be described by setting, for example, $\langle\Phi_3 \rangle =v^{3/2}$ everywhere, magnetic charges involve field configurations of $\Phi_a$ which cannot be made globally homogeneous by smooth gauge transformation. The appropriate $SU(2)$ invariant expression for the embedded $U(1)_{\rm ax}$ field strength is known as the 't Hooft tensor \cite{tHooft:1974kcl}
\begin{align}
\label{eq:tHooft_tensor}
         \mathcal{F}_{\mu\nu} &= \hat{\Phi}_a F^a_{\mu\nu} -  {1 \over g'_5}\varepsilon^{abc}\hat{\Phi}_a \mathcal{D}_\mu \hat{\Phi}_b \mathcal{D}_\nu \hat{\Phi}_c \ , 
\end{align}
where $\hat{\Phi}_a \equiv \Phi_a/|\Phi|$, and $g'_5$ is the non-Abelian gauge coupling. The magnetic flux through a closed Gaussian surface $\Sigma$ is now directly related to the topological degree of the map $\hat{\Phi}_a$ from $\Sigma$ to $S^2 \simeq SU(2)/U(1)_{\rm ax}$,
\begin{align}
\label{eq:BfluxIsTopologicalDegree}
    \int_\Sigma dS^{ij}  \mathcal{F}_{ij}
    =  {1 \over g_5'} \int_{\Sigma} dS^{ij}\varepsilon^{abc} \hat{\Phi}_a\partial_i\hat{\Phi}_b\partial_j\hat{\Phi}_c   = {2\pi n \over g_5}\ ,
\end{align}
where $n\in \mathbb{Z}$ and the $U(1)_{\rm ax}$ gauge coupling $g_5=g'_5/2$ is rescaled so that the smallest possible electric charge in the theory is unity (descending from the fundamental or other half-integer representations of $SU(2)$). 

In the vicinity of the core, the configuration will look spherically symmetric as in $\mathbb{R}^3$, we can take $\Sigma=S^2$, and one can recover the familiar `hedgehog' behavior $\Phi_a \propto  r \hat{r}^a + \mathcal{O}(r^3)$ (along with $A^a_i \propto r \epsilon_{aij}\hat{r}^j+ \mathcal{O}(r^3)$).
$\mathcal{F}$ satisfies the Bianchi identity $d\mathcal{F}=0$ everywhere, except at a symmetry restoration point $|\Phi|=0$, where it is not defined, and the magnetic core is smoothly resolved by a region trapped in the unbroken $SU(2)$ phase. The effective size of this non-Abelian core is controlled by the mass of the $W$ gauge boson in the broken phase $\ell^{-1}_{\rm core}\sim M_W = g'_5v^{3/2}$.

\begin{figure}[t]
        \centering
         \includegraphics[width=0.4\textwidth]{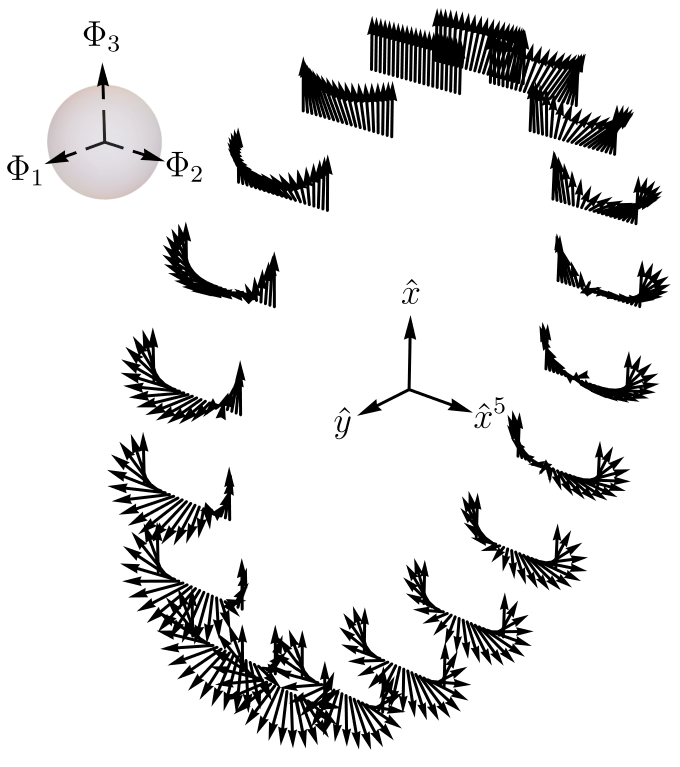}
         %
\caption{
Higgs winding at spatial infinity for a unit-magnetic-charge 
't Hooft--Polyakov configuration on \(\mathbb{R}^2\times S^1\).
At large radius \(\sqrt{x^2+y^2}\), the boundary is a torus: 
\(\theta\) is the visible circular direction, while \(x^5\) runs across 
the strip with periodicity implicit. The arrows show the  adjoint 
Higgs field \(\hat\Phi_a\), oriented with respect to the target-space 
\(S^2\) (top left), defining a degree-one map \(T^2\to S^2\). See 
\cref{eq:Degree1HiggsMap} for details.
}
        
        %
        \label{fig:HiggsWinding}
\end{figure}

At distances of order or larger than $L$, the configuration is significantly distorted from the $\mathbb{R}^3$ solution. The natural Gaussian surface becomes more like a torus $\Sigma=T^2$, which is indeed the topology of infinity in $\mathbb{R}^2\times S^1$.  Just like maps from $S^2$ to $S^2$, maps from $T^2$ to $S^2$ can have non-trivial wrapping; their homotopy class is classified by an integer degree $n$, which determines the magnetic charge as per \cref{eq:BfluxIsTopologicalDegree}. 
Although not crucial for our purposes,  it is easy to construct an explicit example $\hat{\Phi}_a$ with $n=1$, as can be found in \cref{app:HiggsProfile}, and plotted in \cref{fig:HiggsWinding}, for the artistic delight of one of the authors.






\subsection{The string tension.}

In the regime of interest $L/\ell_{\rm core} \sim M_W L \gg 1$, the string tension $\mu$ can be split cleanly between UV and far-field contributions. From the 4D perspective, the latter, log-divergent contribution comes from gradient energy in the wound axion field.
From the 5D perspective it follows from the focusing of magnetic field lines, as $|\vec{B}|^2 \rightarrow 1/(L\rho g_5)^2$ for $\rho =\sqrt{x^2+y^2}\gg L$, contributing  $(\pi  /g_4^2L^2)\ln(\rho/L)$ to $\mu$ (focusing now on $n=1$). An extra UV contribution instead comes from the tension of the magnetic string core, which can be taken the same as in $\mathbb{R}^4$. This is bounded below by the result $4\pi M_W/g_5'^2$ for the BPS case ($\lambda = 0$),
and increases very modestly away from it \cite{tHooft:1974kcl,Forgacs:2005vx}. Adding both contributions and rewriting them in terms of $f$, 
\begin{align}
\label{eq:string_tension}
   \mu \approx \pi f^2 \left({M_W L} + \ln\left( \rho/L \right) \right) \ .
\end{align}
As usual,  the IR divergence will be cut off at the inter-string separation. 

\subsection{Axion quality.}

A 4D axion descending from a $U(1)$ gauge field in 5D receives contributions to its potential from electrically charged bulk fields. In the $SU(2)\rightarrow U(1)_{\rm ax}$ construction above, there will thus be an unavoidable contribution from the massive $W$  gauge bosons. The one loop potential can be evaluated exactly using expressions for example in ref.~\cite{Arkani-Hamed:2007ryu} (see \cref{eq:V_W_general}). In the limit of interest,
\begin{align}
        \label{eq:WbreakingPotential}
V_W &\simeq  { M_W^2 \over  \pi^2 L^2}e^{-M_W L }\cos(2a/f) \ , \quad \text{for } \ M_WL \gg 1 \, .
\end{align}
Notice $V_W$ breaks the axion shift symmetry down to a $\mathbb{Z}_2$ that is preserved, due to $W$ carrying charge 2 with respect to the unbroken $U(1)_{\rm ax}$. This $\mathbb{Z}_2$ is broken by states of unit charge, which in turn come from half-integer representations of the original $SU(2)$.

$V_W$ is exponentially suppressed, but it cannot be made arbitrarily small, for fixed $g'_5$. This is because we want $M_W$, as well as $v$, and the string core energy density $\rho_{\rm str}\approx \mu M^3_W$, to all lie below the cut-off scale $\epsilon \Lambda_5$, where $\Lambda_5 \sim 24\pi^3/ g_5'^2 C_2 $ is the NDA strong coupling scale of non-Abelian gauge theory in 5D (with quadratic Casimir $C_2=N$ for $SU(N)$), and $\epsilon < 1$ determines how weakly coupled is the UV completion. 
These three relevant scales do not have a fixed relative hierarchy, as shown in \cref{fig:FinalPlot}, but the most permissive case of $\epsilon =1$, gives a hard lower bound on the axion mass $m_a$, roughly $m_a^2 \gtrsim f^2 e^{-12\pi^3 /g_4'^2}$. 

We note finally that, interpreting $M_WL$ as the leading instanton action, as suggested by \cref{eq:WbreakingPotential}, the result for the string tension \cref{eq:string_tension} is parametrically consistent with that of ref.~\cite{Reece:2024wrn} for stringy axiverse strings.



\subsection{The QCD Axion}
In 4D the QCD axion is defined by its coupling to the QCD topological charge density 
$a \, \varepsilon^{\mu\nu\rho\sigma} G_{\mu\nu}G_{\rho\sigma}$, where $G$ is the field strength of the $SU(3)_c$ of QCD. For an axion descending from $U(1)_{\rm ax}$ in 5D, this descends from a mixed Chern-Simons (CS) term $\varepsilon^{MNRST}A_M G_{NR} G_{ST}$. However, having embedded $A_M \rightarrow A^a_M$ into $SU(2)$ to resolve axion strings, the latter does not appear manifestly gauge invariant. 
The apparent obstruction is easily overcome.  We can, for example, start from a unified gauge group 
 $\mathcal{G}$ that spontaneously breaks to $\mathcal{H}\supset SU(3)_c \times U(1)_{\rm ax}$. As long as $\pi_2(\mathcal{G}/\mathcal{H})\neq 0$, the theory will contain smooth 't Hooft-Polyakov-like configurations carrying magnetic charge for $U(1)_{\rm ax}$ (see e.g. ref.~\cite{Weinberg:1992hc}  for a review). The desired physics then emerges from the gauge-invariant non-Abelian CS term for $\mathcal{G}$
\begin{align}
\label{eq:Non-Abelian_CS}
\begin{split}
    S_{\rm CS} =& \frac{k g_5'^3}{96\pi^2}\int d^5x \, \varepsilon^{MNRST} \, {\rm Tr}\Big(
 \tilde{A}_M \tilde{F}_{NR} \tilde{F}_{ST} \\
 & - \tilde{A}_M \tilde{A}_N \tilde{A}_R \tilde{F}_{ST} 
 + \frac{2}{5} \tilde{A}_M \tilde{A}_N \tilde{A}_R \tilde{A}_S \tilde{A}_T \Big) \ ,
\end{split}
\end{align}
where $g_5'$ is now the unified gauge coupling, $\tilde{F}_{MN}=\tilde{F}_{MN}^a\tau^a$ its field strength and $\tilde{A}_M=\tilde{A}_M^a\tau^a$ its gauge potential, with $\tau^a$ the generators of $\mathcal{G}$. 
Upon breaking $\mathcal{G}\rightarrow SU(3)_c \times U(1)_{\rm ax} \times ...$, and decomposing $\tilde{A}^a \tau^a = \tilde{A}^i\tau^i + A \tau^{\rm ax} + \dots$, with $i$ running over QCD generators and $\tau^{\rm ax}$ the $U(1)_{\rm ax}$ direction, \cref{eq:Non-Abelian_CS} decomposes, and reveals the mixed CS term desired,
\begin{align}
\label{eq:CSdecomposition}
    \mathcal{S}_{\rm CS} \supset {g_5 g_{s,5}^2 N_w \over 32 \pi^2}\int d^5x  \, \varepsilon^{MNRST} \, A_M {\rm Tr}\left(
  G_{NR} G_{ST} \right) \ ,
\end{align}
where $g_{s,5}$ is the 5D strong coupling.
The coefficients $k,N_w\in \mathbb{Z}$ in \cref{eq:Non-Abelian_CS,eq:CSdecomposition} are necessarily quantized to ensure invariance under large gauge transformations. $N_w$ ultimately controls the minimum number of domain walls attached to a QCD axion string. Whether we can have $N_w=1$ depends on $\mathcal{G}$ and the breaking pattern, but it is easy to construct simple examples where $k=1$ descends to $N_w=1$, such as $\mathcal{G}=SU(4)\rightarrow SU(3)_c\times U(1)_{\rm ax}$, with $\tau^{\rm ax} = $ diag$(1,1,1,-3)/2\sqrt{6}$. Further details of this can be found in \cref{app:SUNmodelDetails}.

A unified group $\mathcal{G}$ is not essential. The mixed CS term \cref{eq:CSdecomposition} can also be induced by fermions charged under both $SU(2)\rightarrow U(1)_{\rm ax}$ and $SU(3)_c$, as also suggested in ref.~\cite{Sehayek:2026pvu}. 
As a simple model, we can add a bifundamental $\psi$ with bare mass $m_\psi$ and a Yukawa interaction $y\bar{\psi}\Phi \psi$. Upon symmetry breaking, $\psi$ decomposes into states with masses $m_\psi \pm y v^{3/2}$ and $U(1)_{\rm ax}$ charge $q=\pm 1$. At one loop, integrating out a single 5D Dirac fermion produces fractional CS coefficients, by themselves not gauge invariant \cite{Redlich:1983kn,Alvarez-Gaume:1984zst,Witten:1996qb,Bonetti:2013ela}. For the mixed $U(1)\times SU(N)$ case of interest, this is a contribution $q\, \text{sign}(m)/2$ to $N_w$, with $m$ the fermion mass \cite{Ohmori:2014kda}. As long as $y v^{3/2} > m_\psi$, the two contributions add rather than cancel, leading to $N_w=1$ (see ref.~\cite{Boyarsky:2002ck} for an analogous  discussion of the $U(1)^3$ CS in an $SU(2)\rightarrow U(1)$ model).

\begin{figure}[t]
        \centering
         \includegraphics[width=0.5\textwidth]{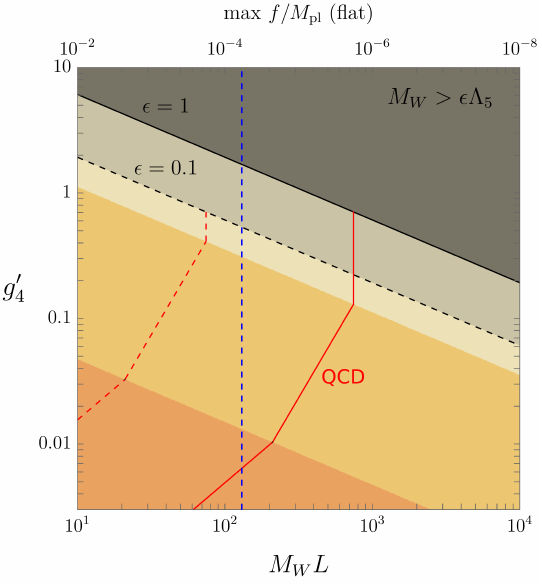}
         %
        \caption{Starting with the minimal $SU(2)\rightarrow U(1)_{\rm ax}$ construction, we represent parameter space in terms of $M_WL$, which controls the axion quality (since irreducible breaking effects scale as $e^{-M_WL}$), and the non-Abelian 4D gauge coupling $g_4'$, which controls the  separation between KK scale  and maximum cut-off of the 5D EFT, $\Lambda_5 L \sim  24 \pi^3/g_4'^{2}C_2$.
        All lines are to be understood as rough boundaries. 
        Different background shades denote regions where (from bottom to top) $v ,  (\rho_{\rm str})^{1 \over 5}, M_W \leq \epsilon \Lambda_5$ are largest. In the most permissive case $\epsilon =1$, only the top gray region is excluded, while a more weakly coupled UV completion $\epsilon=0.1$ is shown for comparison.
        For our models implementing the QCD axion, the presence of QCD in the bulk sets a fixed upper bound on the 5D cut-off. Demanding $M_W ,  (\rho_{\rm str})^{1 \over 5}, v$ lie below it, further excludes regions to the right of the bold red line (red dashed for $\epsilon=0.1$  instead).
        A similar bound will apply also to ALPs coupled to photons with electroweak gauge groups in the bulk.
        The vertical dashed blue line roughly marks the minimum quality required to solve the strong CP problem. 
        }
        %
        \label{fig:FinalPlot}
\end{figure}

For the QCD axion to solve the Strong CP problem, the unavoidable contribution to its potential from integrating out the charged gauge bosons should be highly subdominant. To account for the unobserved neutron EDM \cite{Crewther:1979pi,Abel:2020pzs}, we require $V_{W}/V_{\rm qcd} \lesssim 10^{-10}$, where $V_{\rm qcd}\sim m_u \Lambda_{\rm qcd}^3$ is controlled by the up-quark mass and the QCD confinement scale, which translates to a non-trivial bound on $M_WL\gg1$.  Roughly,  $M_W L \gtrsim 110-180 \quad \text{for} \quad f \sim  10^{8\div 16}\GeV$. 
This however, as highlighted in \cref{fig:FinalPlot}, is not far from the upper bound obtained by demanding $M_W$ (as well as $v,\rho_{\rm str}^{1/5}$) stay within the validity of the 5D EFT, whose cut-off has now a fixed upper bound from the presence of $SU(3)_c$ in the bulk,
 $ \epsilon \Lambda_5 L \lesssim \epsilon 250 /g_{s,4}^2 $, with $g_{s,4}=g_{s,5}/\sqrt{L}\sim 0.5$ the 4D QCD coupling at the KK scale.

\subsection{Axion-like particles (ALPs)}
\label{sec:Photons}

Beyond the QCD axion, axiverse axions are of particular experimental interest when coupled to photons through ${1\over 8}g_{a\gamma\gamma}\,a\,\epsilon^{\mu\nu\rho\sigma}f_{\mu\nu}f_{\rho\sigma}$, conventionally normalized as $g_{a\gamma\gamma}=e^2 c_\gamma/ 8\pi^2 f $, where
$f_{\mu\nu}$ is the electromagnetic field strength, $e$ its gauge coupling, and $c_\gamma$ a model-dependent coefficient.
Again, we can implement this in two ways, mirroring the previous discussion.
We can start from a unified gauge group $\mathcal{G}\rightarrow \mathcal{H}$, with $\mathcal{H}$ now containing $U(1)_{\rm ax}$ and one (or both) of the Standard Model (SM) electroweak factors $SU(2)_{L}\times U(1)_Y$, and write down a non-Abelian CS term \cref{eq:Non-Abelian_CS} for $\mathcal{G}$. 
A simple example, which avoids also producing magnetic monopoles for $U(1)_Y$, is $\mathcal{G}=SU(3)\rightarrow SU(2)_{L}\times U(1)_{\rm ax}$, with $\tau^{\rm ax} = $ diag$(1,1,-2)/2\sqrt{3}$. 
The resulting mixed $U(1)_{\rm ax}\times SU(2)_{L}^2$ CS term in 4D gives an axion coupling to the $SU(2)_{L}$ topological charge density, which further decomposes upon 
electroweak symmetry breaking, to give the desired axion-photon coupling with $c_\gamma = k$.

Alternatively, we may add to the minimal construction a fermion in the fundamental representation of $SU(2)\rightarrow U(1)_{\rm ax}$, also charged under the electroweak sector, for example $U(1)_Y$ with charge $q_Y$. As before, as long as it gets most of its mass from $\langle \Phi \rangle$, integrating out the fermion produces a CS level $q_Y^2$ for the mixed $U(1)_{\rm ax} \times U(1)_Y^2$ term, and after descending to 4D and electroweak symmetry breaking, gives $c_\gamma = 2q_Y^2$.

\subsection{More general constructions}
\label{sec:Orbifolding}

\paragraph{Orbifolds.}

We have so far focused on an $S^1$ compactification for simplicity, but a more phenomenological 5D model (e.g. that allows for 4D chiral fermions) will involve a segment $x^5\in [-{L/ 2},{L/ 2}]$ with boundary conditions (b.c.) removing certain 4D degrees of freedom. 
The rules of consistent b.c. are found for example in ref.~\cite{Csaki:2003dt}.
It is natural to expect that b.c. retaining the 4D axion, such as those specified below, are also compatible with 5D magnetic configurations that appear as axion strings in 4D, and we find this is indeed the case. Moreover, the same b.c. act as a repulsive force on the string core and forbid magnetic strings from ending on a boundary, thus ensuring the stability of strings arbitrarily longer than $L$.

In the Dirac limit, we only have the $U(1)_{\rm ax}$ gauge field. Standard 
b.c. $A_\mu  , \, \partial_5 A_5 = 0$ at $x^5=\pm L/2$ (corresponding, in an $S^1/\mathbb{Z}_2$ orbifold construction, to odd and even parity for $A_\mu$ and $A_5$ respectively) preserve the 4D axion, and remove the 4D gauge field. Compatibility can be explicitly checked against the construction in \cref{eq:DiracAxionString}. 

For the minimal $SU(2)\rightarrow U(1)_{\rm ax}$ model, orbifold parity assignments consistent with a 4D axion are $A^{1}_\mu,A^{2,3}_5, \Phi_{2,3}$ even, and $A^{1}_5,A^{2,3}_\mu,\Phi_1$ odd. So are the rank-reducing b.c., Neumann on $A^{a}_5,\Phi_3$ and Dirichlet on $A_\mu^{a},\Phi_{1,2}$ for $a=1,2,3$. Both are compatible with the explicit $\Phi$ configuration of \cref{eq:Degree1HiggsMap} plotted in \cref{fig:HiggsWinding}.

Above we presented constructions implementing the QCD axion, or axion-photon couplings, of the form $SU(N)\rightarrow SU(N-1)\times U(1)$, with $N=2,3$. Clearly we require Neumann b.c. for the $A^a_\mu$ of the $SU(N-1)$ subgroup to retain QCD and weak interactions, for the $A_5$ of $U(1)_{\rm ax}$ to retain the axion, and for $\Phi_3$ to implement $\langle \Phi_3\rangle = v^{3/2}$. It can be easily shown that consistency then requires Dirichlet for all the $A^a_\mu,\Phi^a$ associated to the mixing of $SU(N-1)$ with $U(1)_{\rm ax}$ directions. 

\paragraph{More dimensions and higher rank.}

We have focused on a 1-form $A_M$ in 5D. Our analysis trivially extends to Wilson-line axions from 1-form gauge fields on non-trivial 1-cycles in higher dimensions; the magnetically charged objects are the corresponding codimension-3 't Hooft-Polyakov solitons (e.g. a magnetic `wall' in 6D), which become 4D axion strings when appropriately compactifying along their worldvolume.


We expect also that a similar analysis should extend to axions descending from periods of higher-rank $p$-form gauge fields over non-trivial $p$-cycles in higher dimensions, although the nature of the soliton may differ.
For example, in the case of a 2-form in 6D, instantons from a gauge theory coupled to the 2-form can serve as magnetically charged solitons.  We defer a proper study to future work.

\section{Cosmology}
\label{sec:Cosmology}

\paragraph{Production.}

The virtue of resolving axion strings within field theory is that their cosmological production can now proceed straightforwardly, for high enough initial temperature, from a symmetry-breaking phase transition in the early universe. This phase transition is intrinsically higher dimensional, corresponding in 5D to the spontaneous breaking of non-Abelian gauge symmetry down to a subgroup that includes the $U(1)_{\rm ax}$ theory yielding the 4D axion.
What are produced are 't Hooft-Polyakov magnetic strings. As discussed, phenomenological boundary conditions will ensure the strings cannot wrap the extra dimension, and in 4D the only topological defects present will be axion strings.

While the Hubble scale for inflation can lie below the KK scale $H_I \ll L^{-1}$, so that 4D inflation is standard, we require the post-inflationary universe starting with maximum temperature $T_{\rm max}\gg L^{-1}$. More precisely, efficient string production requires thermal energy density in the $U(1)_{\rm ax}$ sector to reach that of the magnetic string core $ T^5_{\rm max}\gtrsim  \rho_{\rm str}$. 
On the other hand, $T_{\rm max}$ should be below some critical value $T_{\rm crit}$ when thermal pressure starts to compete with the stabilization mechanism for the size of the extra dimension, less it decompactify. 
To be calculable, all of this should occur within the 5D EFT, at scales below its cut-off $ \epsilon \Lambda_5 \leq M_5$, where $M_5$ is the 5D Planck scale. In summary, the necessary conditions for the post-inflationary scenario are 
\begin{align}
   \begin{split}
   \label{eq:inequalities}
        &H_I \ll L^{-1} \ll M_W ,  \ \\
     L^{-1} \ll  (\rho_{\rm str})^{1 \over 5} & \lesssim T_{\rm max} \lesssim  T_{\rm crit} \lesssim \epsilon\Lambda_{5}{\leq} M_5 \, ,
   \end{split}
\end{align}
which can, in principle, be met.
It is important to emphasize that in models discussed above implementing the QCD axion and axion-photon couplings, the presence of SM interactions in the bulk sets a fixed maximum scale separation between the KK scale and 5D cut-off $\Lambda_5 L \sim 24 \pi^3/g_{\rm SM}^2C_2$, with $g_{\rm SM}$ the relevant coupling. For a consistent post-inflationary scenario, completely calculable within the EFT, the chain of inequalities in \cref{eq:inequalities} must lie within this window. For QCD axion models described here, the constraint from quality leaves some but not much room to make this work, as seen in \cref{fig:FinalPlot}.

We have focused on a flat extra dimension. This presumes $\sqrt{T_{\rm crit}^5/M_5^3}  \ll L^{-1} $, since $T_{\rm crit}^5$ is the energy density associated to the stabilizing potential. This turns into a condition $M_5^3/L^2 \gg T_{\rm crit}^5 \gtrsim \rho_{\rm str} $ which can be expressed as an upper bound on the axion decay constant $f \ll \mpl /(M_WL)^2$ that we include in the top axis of \cref{fig:FinalPlot}, with $\mpl = \sqrt{M^3_5 L}$ the 4D Planck scale. This same condition guarantees  $L^{-1} \gg  H_I \gtrsim \sqrt{\rho_{\rm str}L/\mpl^2} $, while imposing $M_W ,(\rho_{\rm str})^{1 \over 5}, v \lesssim M_5$ gives less stringent bounds on $f$ everywhere in \cref{fig:FinalPlot}.
We leave the generalisation to a warped extra dimension to future studies. 


\paragraph{{Some phenomenological comments.}}

Cosmic string networks evolve toward an attractor ‘scaling’ solution (see e.g. ref.~\cite{Vilenkin:2000jqa} and references therein) with a near-constant dimensionless string length density. 
The core resolution described here may significantly modify scaling parameters and phenomenology relative to purely 4D models. Beyond the additional contribution to the tension in \cref{eq:string_tension}, localization in the fifth dimension may plausibly reduce the effective intercommutation rate, as strings that intersect in 4D may miss in the extra dimension. This feature is shared with stringy axiverse strings \cite{Copeland:2003bj,Jackson:2004zg}, for which some phenomenological studies have been attempted, e.g. in refs.~\cite{Avgoustidis:2005nv,Pourtsidou:2010gu}. Given the  nature of the resolution, in our case this effect is in principle now fully calculable in field theory. 


The ultimate fate of the string network is set by whether the leading axion potential preserves a discrete symmetry, leaving $N_w$ domain walls attached to each string. For an ALP, the unavoidable $W$ contribution, e.g.~\cref{eq:WbreakingPotential}, may well dominate, with $N_w>1$ given by the $U(1)_{\rm ax}$ charge of the $W$. The resulting string-wall network can then be long-lived, its decay controlled by subleading $\mathbb{Z}_{N_w}$ breaking effects from unit-charged states with mass $M_F>M_W$. If instead these states dominate, $M_F\lesssim M_W$, then $N_w=1$ and the network collapses at $t\sim 1/m_a$, when walls enter the horizon. For the QCD axion, $N_w$ is fixed by the mixed CS term in \cref{eq:CSdecomposition}; as discussed above, we gave explicit models with $N_w=1$.

\section{Axiverse strings dissolved}
\label{sec:Dissolution}

\begin{figure}[t]
        \centering
         \includegraphics[width=0.52\textwidth]{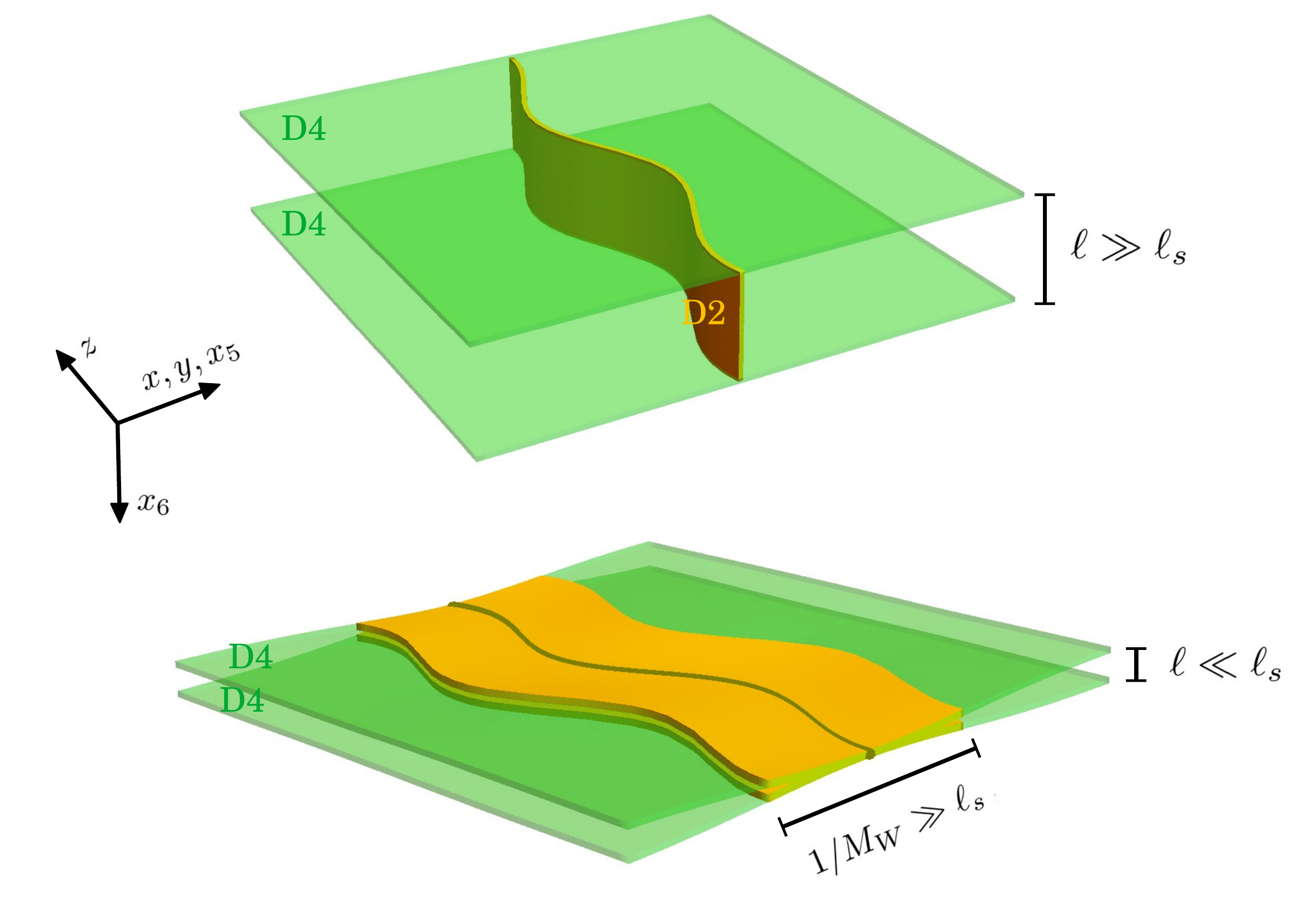}
         %
        \caption{
    Two D4 branes defining a worldvolume 5D gauge theory in $(t,x,y,z,x_5)$. This simple example  shows the object resolving an axion string smoothly going from an elementary D2-brane description to a field-theoretic ’t Hooft–Polyakov configuration, as the D4-brane separation modulus $\ell$ varies from superstringy (top) to substringy (bottom).
        }
        %
        \label{fig:D4D2}
\end{figure}

Our discussions above have been self-contained within (extra-dimensional) field theory. Naively,  axion strings from the original string theory axiverse look very different to the  't Hooft-Polyakov strings discussed so far,
as they typically originate from elementary (stringy) D-brane objects with a tension beyond
the string scale and are associated with axions descending from Ramond-Ramond (RR) $p$-form gauge fields that are intrinsically Abelian.
Despite these apparent differences, we will now show that the  field theory, solitonic solutions do describe
axiverse strings in some part of the moduli space.

We start by embedding
the 't Hooft-Polyakov string solution in a string theory construction, along the lines of ref.~\cite{Diaconescu:1996rk}.
Consider two coincident D4 branes in type IIA string theory. The open string excitations ending
on them describe a $U(2)$ gauge theory in 5D with scalars in the adjoint
representation and a gauge coupling squared $g_5'^2=2 (2\pi)^2 g_s \ell_s$, where $g_s$ and $\ell_s$ are the string coupling and length respectively. 
Separating the two D4 branes by a distance $\ell$ corresponds to giving a vev  to
the adjoint scalar and breaking $U(2)$ to $U(1)\times U(1)$ (each factor living on their respective brane).
For our purposes, we can just focus on the $SU(2)\subset U(2)$, which breaks to  $U(1)_{\rm ax}\subset SU(2)$, identified with the axion-containing factor of previous sections.
The $W$ bosons correspond to open strings stretching
between the branes and have mass $M_W=\ell/(2\pi \ell_s^2)$. As long as $\ell\ll\ell_s$, the breaking
of $SU(2)$ is within the EFT regime, and so are 't Hooft-Polyakov string
solutions charged under $U(1)_{\rm ax}$, which have a macroscopic core $1/M_W \gg \ell_s$ and a low  tension $4\pi M_W/g'^2_5=\ell/(2g_s (2\pi\ell_s)^3)$.
As the D4 brane separation grows past the string size,  the breaking of $SU(2)$ exits the EFT description,
the solitonic string thins below the string scale, and its tension becomes superstringy.
In this regime, the corresponding magnetic charge for $U(1)_{\rm ax}$ is instead resolved by an elementary D2 brane stretching between the two D4 branes.~\footnote{More precisely, a D2 brane ending on a D4 brane acts as a unit magnetic charge for the $U(1)$ living thereon. Stretching from one D4 to the other, it has the right combination to match the unit $U(1)_{\rm ax}$ charge of the 't Hooft-Polyakov solution.} 
Viewed as a string, its tension 
$\ell/(2g_s (2\pi\ell_s)^3)$  matches exactly the one computed before. 
This simple example, depicted graphically in \cref{fig:D4D2}, shows how an elementary stringy D-brane axionic string `dissolves' into a field-theoretic solitonic string as we move in moduli space. 
We note, however, that the axion here arises from the internal component of a vector living on the D4 branes, hence from the open string sector of the theory, unlike the original axiverse.

\begin{figure}[t]
         \includegraphics[width=0.4\textwidth]{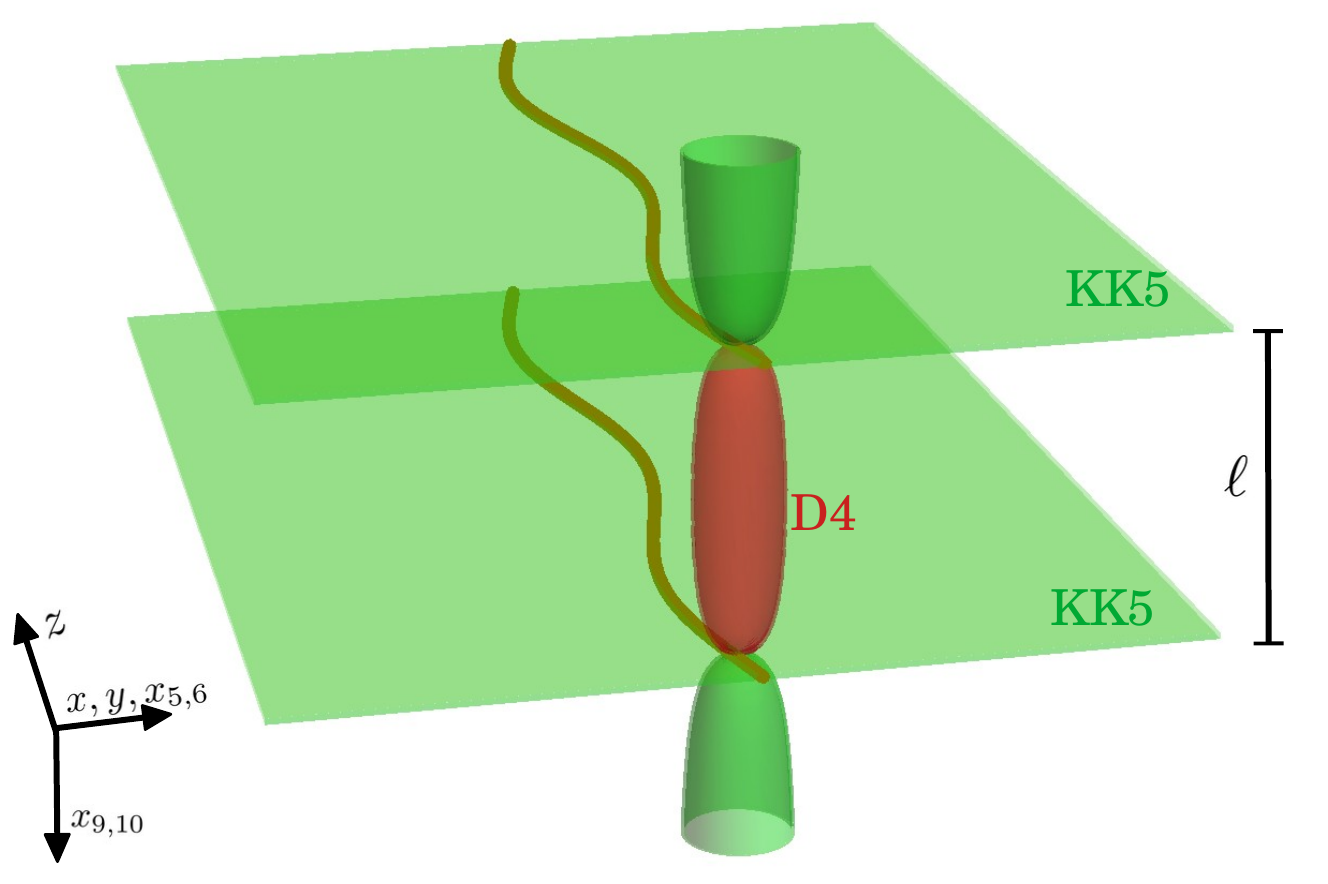}
         %
        \caption{
        Two KK5 monopoles defining a worldvolume 6D gauge theory $(t,x,y,z,x_{5,6})$. 
        Each monopole is a gravitational soliton with a transverse geometry locally resembling a `cigar'. Along the transverse line connecting two monopoles, the cigars meet to form an $S^2$ in between. For superstringy separation $\ell$, the axion string core is resolved by an elementary D4 brane wrapping said $S^2$ as in the figure. For substringy separation this `dissolves' into a field theory solitonic description similar to the bottom of \cref{fig:D4D2}.
        }
        %
        \label{fig:KK5}
\end{figure}

We can  easily find an example
where the axion comes from the RR sector by simply applying a chain of string dualities, starting from the previous D4/D2 brane construction.~\footnote{
One such chain is a T-dualization along two directions orthogonal to the D4/D2-branes, yielding a D6/D4 system, followed by a weak-strong-weak duality that blows up the M-theory circle and then shrinks another direction along the D6 worldvolume.}
The two D4 branes are replaced by two Kaluza-Klein monopoles (KK5), as we attempt to depict in \cref{fig:KK5}. As before, when separated, each individually supports a localized $U(1)$ gauge field, now arising 
from the RR 3-form $C^{(3)}$ decomposed over the harmonic 2-form of the KK5 background. When the two KK5 are coincident, the $U(1) \times U(1)$ gauge theory is again uplifted to $U(2)$, and their separation $\ell$ plays the role of the symmetry-breaking Higgs vev.
Since KK5 has codimension four, the gauge theory under consideration is now 6D, with gauge coupling squared $g_6'^2=2 (2\pi)^3 \ell_s^2$. 
The $W$ bosons required for the
upgrade now come from D2 branes (which couple electrically to $C^{(3)}$) wrapping the 2-sphere $S^2$ living in between the two KK5, and have mass 
$M_W=2\pi {\cal A}_{S^2}/(g_s (2\pi \ell_s)^3)$, with ${\cal A}_{S^2}\propto \ell$ the $S^2$ area.
When the two KK5 coincide, the $S^2$ collapses to zero size ${\cal A}_{S^2}\rightarrow 0$, we get an $A_1$ orbifold singularity, and the wrapped D2-branes (i.e. the W bosons) become massless.
If ${\cal A}_{S^2}$ is non-zero but parametrically smaller than the string scale, the $SU(2)\subset U(2)$ symmetry breaking is within the EFT description, and the modulus controlling the size of the $S^2$ appears as a Higgs in the adjoint representation
(its off-diagonal charged states also coming from wrapped D2 branes). In this regime, we have 't Hooft-Polyakov wall solutions within the EFT, charged under $U(1)_{\rm ax}\subset SU(2)$, with tension $4\pi M_W/g_6'^2=2\pi {\cal A}_{S^2}/(g_s (2\pi \ell_s)^5)$. These walls become strings once we compactify the 6D theory on the worldvolume of the KK5 monopoles to 5D.
As before, we now contrast this with the opposite regime. As the separation $\ell$ between the KK5 grows large in string units, the $W$ states become heavy and exit the EFT description, along with the 't Hooft-Polyakov solution. Now the appropriate object will instead come from an elementary
D4 brane (which carries magnetic charge w.r.t. $C^{(3)}$ and will thus resolve the axion string core) wrapped over the $S^2$. Viewed as a magnetic wall, its tension matches exactly  that of the solitonic one above, which is indeed superstringy for large ${\cal A}_{S^2}$. 
We note finally that, in this regime, contributions to the axion potential come from Euclidean D2 branes wrapped over $S^1\times S^2$ (i.e. the manifold over which $C^{(3)}$ provides the 4D axion), and scale as $e^{-S_{E2}}\sim e^{-M_W L}$, i.e. exactly like the Casimir contribution in \cref{eq:WbreakingPotential} from the light $W$ in the EFT description.

We end this section  by noting that the configuration of two KK5 monopoles considered above also provides an effective description of $K3$ compactifications near the orbifold limit \cite{Sen:1997kz}, and can be generalized to many other geometric constructions, such as CY compactifications near ADE symmetry-enhancement singularities. In these regions, some D-brane configurations become light, enhancing RR gauge fields to approximate non-Abelian groups and `dissolving' stringy objects, such as axiverse strings, into EFT solitons. In this setup, the CS terms required to couple axions to other gauge theories, such as QCD, are automatically present if those gauge-theory branes wrap the same cycles. For example, in the setup above, the worldvolume action of a stack of D6-branes wrapping $S^1 \times S^2$ and describing QCD would automatically contain the coupling
$\int C^{(3)}\wedge G \wedge G$.

\section{Conclusions}

There is  no obstruction to understanding axion cosmic strings for axiverse axions purely within (extra-dimensional) field theory. They are merely magnetic sources for the associated Abelian gauge fields they descend from, and can be resolved by solitonic configurations, such as the 't Hooft-Polyakov solution amply discussed here. 
In string theory, we argued this is the appropriate description in limits of moduli space when elementary D branes `dissolve' into field theoretical solitons. 

Our work shows that there is also, in general, no obstruction to realizing a calculable `post-inflationary' cosmological scenario for axiverse axions, for high enough reheat temperature, via a symmetry breaking phase transition. There remains perhaps a question of plausibility in the case of greatest interest, the QCD axion, since, in simple models studied here, the extra quality constraint pushes the scales of non-Abelian resolution and reheat temperature to cluster towards the EFT cut-off.

We emphasize that this work focused on the original (closed string) axiverse paradigm; axions reside entirely in bulk gauge fields and $f$ is proportional to a KK scale. This is to be distinguished from the more recent `open string axiverse' \cite{Petrossian-Byrne:2025mto,Loladze:2025uvf}, where axions are mostly identified with charged fields localized in the extra dimensions, and $f$ can be well below the KK scale while preserving the same exponential quality.~\footnote{See also refs.~\cite{Berenstein:2012eg,Honecker:2013mya,Cicoli:2013cha,Allahverdi:2014ppa,Choi:2014uaa} for earlier discussions of axions from the open string sector.} In that case, axion strings and the post-inflationary scenario can enjoy a purely 4D description, including for the QCD axion. They can thus also share all phenomenological predictions with 4D models, unlike the closed string axiverse.

Going forward, we see two main avenues to develop. 
On the phenomenological side, it is important to determine any consequences for axion string networks, their evolution and signals, in light of the extra-dimensional core resolution discussed here, for example along the lines suggested in \cref{sec:Cosmology}.
On the theory side, it would be interesting to explore more field-theoretic solitons resolving axion strings in the case of more general higher-rank gauge fields characteristic of the wider axiverse.

\vspace{0.7cm}
\textbf{Note added.}
Throughout the course of this research we have been aware of parallel and independent work by Nathaniel Craig and Amalia Madden, to appear simultaneously with this manuscript.

\subsection*{Acknowledgments}
We thank Bobby Acharya, Nathaniel Craig, Amalia Madden, John March-Russell, Pierluigi Niro and Marco Serone for useful discussions.
GV acknowledges
this research was supported in part by grant NSF PHY-2309135 to the Kavli Institute for Theoretical Physics (KITP).




\appendix

\section{Asymptotic Higgs behavior}
\label{app:HiggsProfile}

Topologically, a degree-one map $T^2\rightarrow S^2$ can be easily obtained by starting with $T^2$ as a square with opposite sides identified, mapping all boundaries to the north pole, and a point inside (e.g. the centre of the square) to the south pole, with  a smooth interpolation in between. An explicit map, with coordinates $\tau_{1,2}\in [-1/2,1/2]^2$ on the unit torus is given by

\begin{equation}
\label{eq:Degree1HiggsMap}
\hat\Phi_a(\theta,x^5)
=
\frac{1}{D}
\begin{pmatrix}
4\tau_1\sqrt{1-2\tau_2^2}\,\bigl(1-q\bigr)\\[5pt]
4\tau_2\sqrt{1-2\tau_1^2}\,\bigl(1-q\bigr)\\[5pt]
q-\bigl(1-q\bigr)^2
\end{pmatrix} \ ,
\end{equation}
where
\begin{align}
D(\tau_1,\tau_2) =
q+(1-q)^2 \ , \\
q(\tau_1,\tau_2) = 4\tau_1^2+4\tau_2^2-16\tau_1^2\tau_2^2 \ ,
\end{align}
and finally, in terms of the physical coordinates,
\begin{align}
    \tau_1= 1/2 - \theta/2\pi  \ ,
\qquad
\tau_2=   x^5/L  \ .
\end{align}
This has degree $n=1$, and therefore supports unit magnetic charge, as can be verified by inspection, plugging it into \cref{eq:BfluxIsTopologicalDegree}. 
While the closed form expression may seem involved, it is a composition of simple steps; a standard stereographic projection of $S^2$ onto the plane  $X_{1,2} \in \mathbb{R}^2$, a compactification from plane to unit disc $X_{1,2}=d_{1,2}/(1-q)$, with $q=d_1^2+d_2^2 \in [0,1]$, and finally from unit disc to unit square $d_{1,2} = 2 \tau_{1,2} (1-2\tau_{2,1}^2)^{1/2}$.

\section{$SU(N)$ model details}
\label{app:SUNmodelDetails}

This appendix contains details for the class of models with breaking pattern $\mathcal{G}=SU(N)\rightarrow  SU(N-1)\times U(1)_{\rm ax}$, for general $N$. In the main text $N=2$ realized the minimal 't Hooft-Polyakov resolution for axion strings, while $N=4$ and $N=3$ were  examples implementing the QCD axion and an ALP-photon coupling respectively (without  the addition of bulk fermions).

We take symmetry breaking  to follow from the vacuum expectation value of a 5D bulk scalar $\Phi$ in the adjoint representation. The vacuum can be described, for example, by taking $\langle \Phi \rangle = v^{3/2}\tau^{\hat{a}}$, with the unbroken $U(1)_{\rm ax}$ along the $\tau^{\hat{a}}=c_N{\rm diag}(1,1,\dots,1-N)$ generator, and $c_N\equiv 1/\sqrt{2N(N-1)}$. Its gauge coupling, rescaled so that the smallest possible charge is unity, is $g_5=c_N g_5'$, while the $SU(N-1)$ unbroken subgroup inherits the original gauge coupling $g'_5$.
The massive $W$ gauge bosons form a fundamental representation of the unbroken $SU(N-1)$, with $U(1)_{\rm ax}$ charge $N$, and mass $M_W = g_5'v^{3/2}N c_N$, where $N c_N$ is roughly constant.

The theory includes 't Hooft-Polyakov like configurations for all $N$ carrying magnetic charge for $U(1)_{\rm ax}$ which project down to axion strings for the associated 4D axion. The core contribution to the string tension $\mu$ can be approximated by the BPS result
\begin{align}
    \mu \supset {4 \pi v^{3/2} \over g'_5} = {4 \pi M_W \over N c_N g_5'^2} \ .
\end{align}
The irreducible  contribution to the axion potential from integrating out the charged W states at one loop can be obtained exactly 
\begin{align}
\label{eq:V_W_general}
    V_{W} &=  \frac{n_W M_W^4}{4\pi^2} \, \text{Re} \left[ 
\frac{3\, \text{Li}_5\left(\omega\right)}{(M_W L)^4} + 
\frac{3\, \text{Li}_4\left(\omega\right)}{(M_W L)^3} + 
\frac{\text{Li}_3\left(\omega\right)}{ (M_W L)^2} 
\right]  \ ,
\end{align}
where $\omega \equiv e^{-M_WL + i N a/f}$ and $n_W=4(N-1)$ just counts the number of charged degrees of freedom. Expanding \cref{eq:V_W_general} and choosing $N=2$, gives the simple form of \cref{eq:WbreakingPotential} quoted in the main text.

For $N>2$, we can include the non-Abelian Chern-Simons term of \cref{eq:Non-Abelian_CS} in the theory. Its decomposition is most easily seen by writing it as a 6D integral \cite{Wu:1983kz}
\begin{align}
 S_{\rm CS} = \frac{k g_5'^3}{24\pi^2}\int_{\chi_6}  \, {\rm Tr}\Big(
 \tilde{F}\wedge \tilde{F} \wedge \tilde{F} \Big) \ ,
\end{align}
where $\chi_6$ is any manifold whose boundary is the 5D theory of interest ($k \in \mathbb{Z}$ ensures all choices are equivalent), and $\tilde{F} = {1\over 2}\tilde{F}_{MN} \, (dx^M \wedge dx^N)$. Upon separating $\tilde{F} = F^{\rm ax}+G \dots$ into its $U(1)_{\rm ax}$ and $SU(N-1)$ components, the mixed term becomes  $3\, {\rm Tr}(F^{\rm ax} \wedge G \wedge G)$, where the factor of 3 is combinatorial. For our specific model, this simplifies to  $3 c_N F^{\rm ax} \wedge {\rm Tr}(G \wedge G)$. Focusing on $dA \subset F^{\rm ax}$, the mixed term is an exact form (since $d(G\wedge G)=0$), its integral over $\chi_6$ becomes a boundary term, which is the 5D term \cref{eq:CSdecomposition} with mixed CS level $N_w=k$, and $g_{s,5}=g'_5$. 

\bibliography{main} 

\end{document}